\documentclass{article}
\usepackage[utf8]{inputenc}
\usepackage{authblk}
\usepackage{setspace}
\usepackage[margin=1.25in]{geometry}
\usepackage{graphicx}
\graphicspath{ {./figures/} }
\usepackage{subcaption}
\usepackage{amsmath}

\usepackage[style=nejm, 
citestyle=numeric-comp,
sorting=none]{biblatex}
\title{Balancing property optimization and constraint satisfaction for constrained multi-property molecular optimization}

\author[1$\dag$]{Xin Xia}
\author[2$\dag$]{Yajie Zhang}
\author[3]{Xiangxiang Zeng}
\author[2]{Xingyi Zhang}
\author[1]{Chunhou Zheng}
\author[1*]{Yansen Su}

\affil[1]{The Key Laboratory of Intelligent Computing and Signal Processing of Ministry of Education, School of Artificial Intelligence, Anhui University, Hefei 230601, China.}
\affil[2]{The Key Laboratory of Intelligent Computing and Signal Processing of Ministry of Education, School of Computer Science and Technology, Anhui University, Hefei 230601, China.}
\affil[3]{College of Computer Science and Electronic Engineering, Hunan University, Changsha, 410012, China.}
\affil[*]{Email: suyansen@ahu.edu.cn}
\affil[$\dag$]{These authors contributed equally to this work.}

\date{}

\begin{document}

\maketitle

\begin{abstract}
Molecular optimization, which aims to discover improved molecules from a vast chemical search space, is a critical step in chemical development.
Various artificial intelligence technologies have demonstrated high effectiveness and efficiency on molecular optimization tasks. 
However, few of these technologies focus on balancing property optimization with constraint satisfaction, making it difficult to obtain high-quality molecules that not only possess desirable properties but also meet various constraints. 
To address this issue, we propose a constrained multi-property molecular optimization framework (CMOMO), which is a flexible and efficient method to simultaneously optimize multiple molecular properties while satisfying several drug-like constraints.
CMOMO improves multiple properties of molecules with constraints based on dynamic cooperative optimization, which dynamically handles the constraints across various scenarios. Besides, CMOMO evaluates multiple properties within discrete chemical spaces cooperatively with the evolution of molecules within an implicit molecular space to guide the evolutionary search.
Experimental results show the superior performance of the proposed CMOMO over five state-of-the-art molecular optimization methods on two benchmark tasks of simultaneously optimizing multiple non-biological activity properties while satisfying two structural constraints. Furthermore, the practical applicability of CMOMO is verified on two practical tasks, where it identified a collection of candidate ligands of $\beta$2-adrenoceptor GPCR and candidate inhibitors of glycogen synthase kinase-3$\beta$ with high properties and under drug-like constraints.
\end{abstract}

\section{Introduction}
Molecular optimization, which aims to improve molecular properties by modifying molecular structures, is a critical step for several engineering applications, such as drug discovery~\cite{Sheng2017Chemical,vamathevan2019applications} and matrial science~\cite{nigam2021beyond}. 
Molecular optimization is a challenging task in drug discovery, since it always needs to optimize multiple properties that may be conflicted with each other~\cite{fromer2023computer,li2023simultaneously}. 
Additionally, the practical optimization of molecules often necessitates the adherence to stringent drug-like criteria, thereby preventing some molecules from becoming drug candidates~\cite{lee2022docking,wang2021multi}. 
For instance, in order to discover potent inhibitors against the discoidin domain receptor 1, Zhavoronkov et al.~\cite{zhavoronkov2019deep} adhered to stringent drug-like criteria, such as avoiding molecules with structural alerts or reactive groups, to select candidates for subsequent synthesis.
Note that, some stringent drug-like criteria for molecular optimization are typically not suitable to be treated as optimization objectives.
Instead, they are more aptly employed as constraints that guide the optimization process.
For example, the molecules with either small rings ($<5$ atoms) or large rings ($>6$ atoms) are difficult to be synthesized (see Figures S1 in the Supplementary Material). Although some molecular optimization tasks consider the ring size in optimization objectives, e.g., the penalized logP~\cite{JT-VAE}, it is still difficult to rule out the molecules with small or large rings. Consequently, the criteria for molecules with precisely $5$ or $6$ atoms in their rings is typically treated as a constraint in molecular optimization~\cite{liu2018constrained, eckmann2022limo}. 
Thus, the modification of a specific molecule necessitates the optimization of its multiple molecular properties while simultaneously adhering to several drug-like criteria, i.e., it is required to achieve the balance between property optimization and constraint satisfaction.

Recently, molecular optimization has witnessed significant advancements by the application of artificial intelligence methods~\cite{sadybekov2023computational,mak2023artificial,ivanenkov2023chemistry42,Mars}, such as evolutionary algorithms (EA)~\cite{GA+D,GEGL}, reinforcement learning (RL)~\cite{REINVENT,RationaleRL}, and deep learning models (DL)~\cite{zeng2022deep, ScaffoldMD,zhang2024domain}.
Most of these existing molecular optimization methods are proposed to enhance a specific property, such as the quantitative estimate of drug-likeness (QED) or the penalized logP value (PlogP).
Although these single-property optimization methods perform well in exploring vast chemical search spaces and reducing costs~\cite{MolSearch, blanco2023role}, they still face challenges when applied to practical molecular optimization tasks that involve simultaneously enhancing multiple conflicting properties~\cite{VJTNN,Core,UGMMT,SCVAE}.
Afterwards, several multi-property optimization methods have been proposed to deal with the enhancement of multiple molecular properties.
For example, QMO~\cite{QMO} and Molfinder~\cite{MolFinder}, which aggregated multiple molecular properties into a single objective for molecule optimization,  have exhibited promising performance in the simultaneous optimization of multiple properties. 
However, the performance of these methods is prone to be affected by the improper setting of weights. 
MOMO~\cite{xia2022molecule} employed a multi-objective optimization strategy and effectively identified a set of diverse and novel molecules to enhance the likelihood of successful multi-property optimization of molecules.

Despite the rapid advancements in multi-property molecular optimization methods, we still face notable challenges when dealing with practical molecular optimization problems, since existing methods often yield new molecules that fail to meet drug-like constraints.
To date, only a few molecular optimization methods have been proposed to deal with constraints by very simple strategies.
For example, MSO~\cite{MSO} aggregates all the properties to be optimized and the predefined constraints into a single fitness function, which encounters the difficulty in terms of parameter tuning.
GB-GA-P~\cite{GB-GA-NSGAII} is a genetic algorithm-based method for multiple property optimization, and it uses a relatively rough strategy to adhere drug-like criteria by discarding infeasible molecules.
Although above work generate some molecules that satisfy the constraints, the quality (e.g., the 
molecular properties) of the resulting molecules remains to be improved due to the lack of a good balance between the property optimization and the constraint satisfaction.

\begin{figure}[!h]
 \centering
\includegraphics[width=0.72\textwidth]{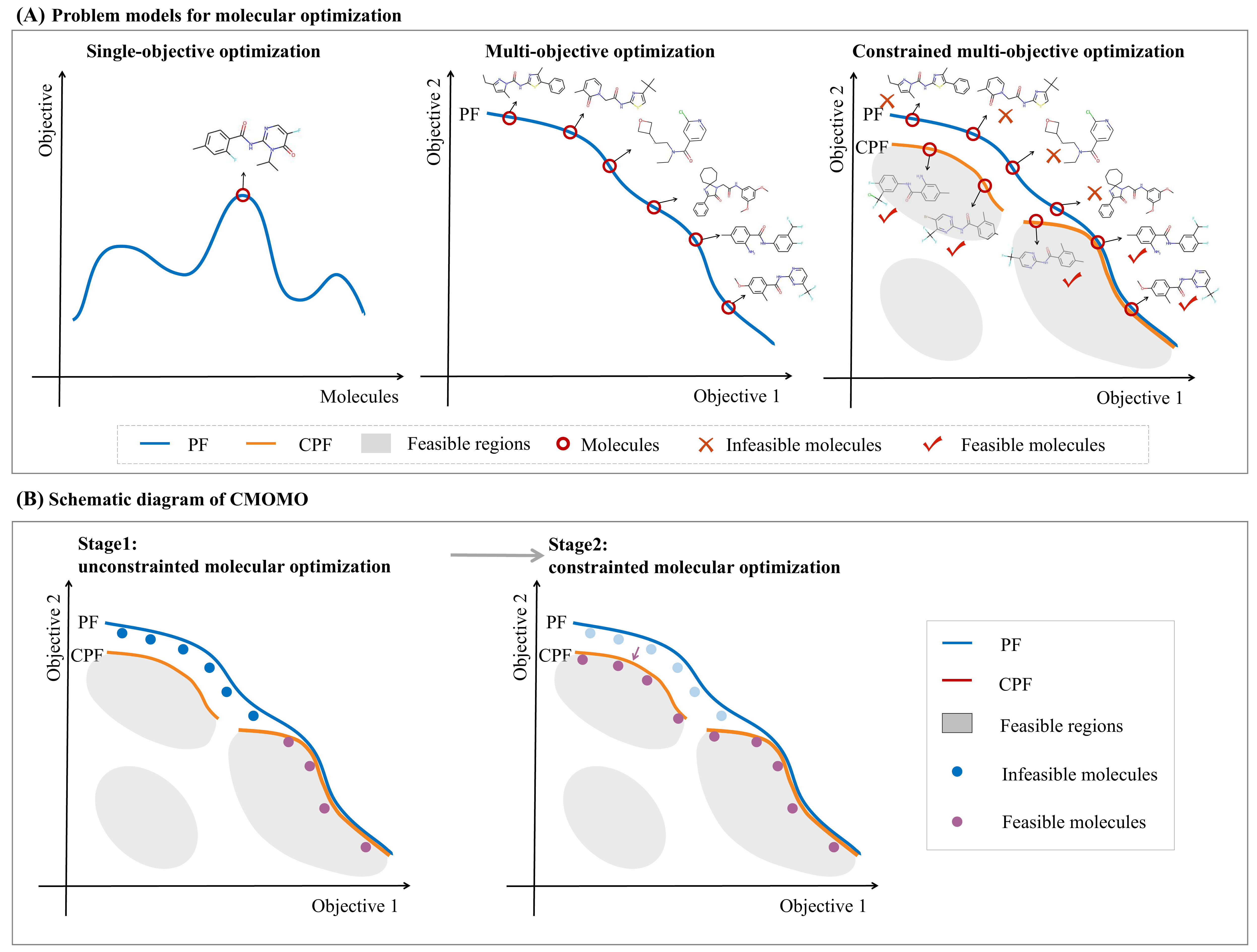}
 \caption{Three problem models for molecular optimization and the schematic diagram of CMOMO. 
 (A) Single-objective optimization aims to find a molecule with the best objective value. 
 Multi-objective optimization aims to search for a set of trade-off molecules among multiple properties in the Pareto front (PF). 
 Constrained multi-objective optimization aims to identify a set of molecules in the constrained Pareto front (CPF) that trade off multiple properties and meet drug-like constraints. 
 (B) The optimization process of CMOMO framework. 
 CMOMO firstly focuses on property optimization to find molecules positioned on the PF, and then discovers the molecules on the CPF to solve the constrained multi-objective molecular optimization problem.}
 \label{problem}
\end{figure}

Given the importance of balancing the property optimization and the constraint satisfaction, the multi-property molecular optimization with constraints is suitably modeled as a constrained multi-objective molecular optimization problem, which is completely different from single-objective optimization (i.e., the optimization of a single property or the aggregation value of multiple properties) and unconstrained multi-objective optimization.
Specifically, the single-objective optimization always discovers a single molecule with the best objective value, and the multi-objective optimization could find a set of molecules with trade-offs among multiple molecular properties (Fig.~\ref{problem}A).
Compared to the two aforementioned, the constrained multi-objective molecule optimization is more 
complex, since it explores molecules that not only compromise different molecular properties but also meet some predefined drug-like constraints. 
What's more, these constraints may result in the narrowness, disconnection and irregularity of feasible molecular space (Fig.~\ref{problem}A).
Therefore, it is challenging to explore feasible molecules (i.e., the molecules adhere constraints) with multiple desired properties.

To address the above problem, in this study we propose a constrained multi-objective molecular optimization framework (CMOMO) to simultaneously optimize multiple molecular properties while satisfying several constraints.
CMOMO firstly solve the unconstrained multi-objective molecular optimization scenario to find molecules with good properties, and then considers both properties and constraints to find feasible molecules that possess promising properties (Fig.~\ref{problem}B).
The main contributions of this paper are summarized as follows.

\begin{itemize}
\item {A constrained multi-objective molecular optimization framework (CMOMO) is suggested to address the multi-property molecular optimization with several drug-like constraints.
The proposed CMOMO achieves a good balance between the optimization of multiple properties and the satisfaction of constrained molecules by first searching for molecules with good convergence and diversity in the unconstrained scenario and then identifying feasible molecules with the desired property values in the constrained scenario.
To the best of our knowledge, CMOMO is the first method that delicately balances property optimization and constraint satisfaction for molecular optimization, and thereby yielding high-quality molecules exhibiting desired molecular properties while adhering rigorously to drug-like constraints.}


\item{
Compared with state-of-the-art methods, CMOMO has the following significant improvements: it facilitates simultaneous optimization of multiple molecular properties while adhering to various drug-like constraints, and subsequently identifies a set of optimal molecules with trade-offs among multiple objectives. 
We demonstrated the high performance of CMOMO in a variety of molecular optimization tasks. Via CMOMO, we identified a collection of potential ligands of $\beta $2-adrenoceptor GPCR receptor (4LDE) and potential inhibitors against glycogen synthase kinase-3 target (gsk3$\beta $) with multiple higher properties while adhering the drug-like constraints.
}
\end{itemize}

\section{Results}
\subsection{The proposed CMOMO framework}
\begin{figure}[!h]
 \centering
\includegraphics[width=0.85\textwidth]{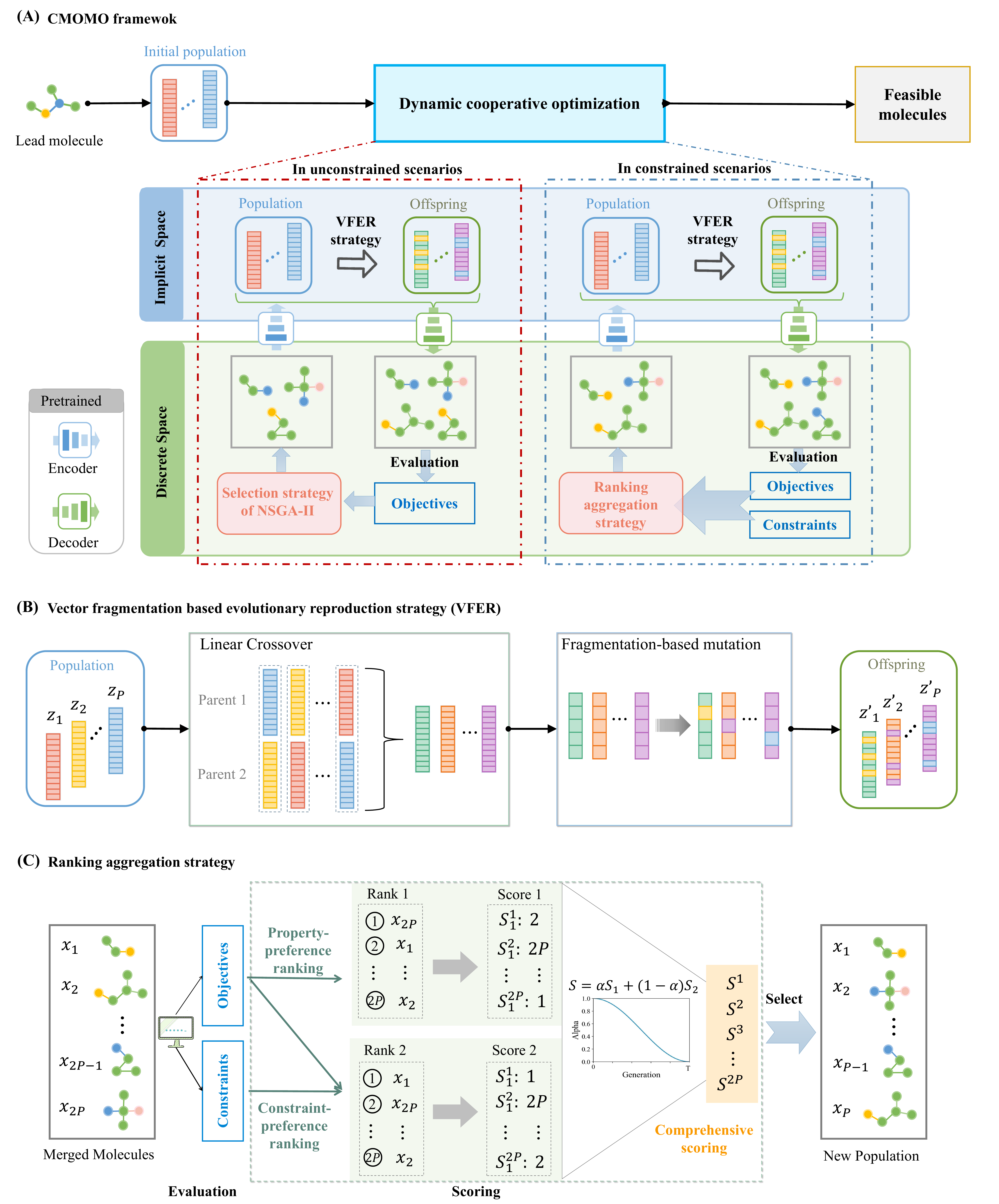}
 \caption{The illustrative diagram of CMOMO. (A) To begin with, CMOMO generates an initial population with $P$ molecules for a lead molecule. 
 Then, CMOMO performs the dynamic cooperative optimization.
 Finally, CMOMO achieves a set of feasible molecules (with desired molecular properties and under the drug-like constraints). 
 (B) The vector fragmentation based evolutionary reproduction strategy (VFER).
 The VFER strategy is employed to generate promising offspring molecules through linear crossover and fragmentation-based mutation operations.
 (C) The ranking aggregation strategy. 
 This strategy dynamically aggregates the rankings of molecules that ordered by properties and constraints, respectively.
 }
 \label{framework}
\end{figure}

In this study, each property to be optimized is treated as an optimization objective and the stringent drug-like criteria are treated as constraints. 
We propose a constrained multi-objective molecular optimization framework CMOMO that includes an efficient population initialization and a dynamic cooperative optimization (Fig.~\ref{framework}A), where the dynamic cooperative optimization serves as the key component.
The driving idea of the dynamic cooperative optimization is to dynamically select high-quality molecules during the evolutionary process and to cooperatively perform the evolution of molecules between the discrete chemical and the continuous implicit spaces.
To this end, a tailored dynamic constraint handling strategy divides the optimization into two scenarios (unconstrained scenario and constrained scenario) and dynamically deals with the constraints in these two scenarios, thereby making the dynamic equilibrium between property optimization and constraint satisfaction. 
Furthermore, a vector fragmentation-based evolutionary reproduction strategy (VFER) significantly enhances the efficiency of evolution in the continuous implicit space.
The above two strategies within the dynamic cooperative optimization help to enhance the ability to identify molecules that possess desirable molecular properties while adhering strictly to drug-like constraints.
The procedure of the proposed CMOMO is illustrated as follows.

(1) \textbf{Population initialization:} Given a lead molecule represented by a SMILES string, CMOMO first utilizes existing public database to construct a Bank library, which contains high property molecules that are similar to the lead molecule. 
Then, CMOMO uses a pre-trained encoder~\cite{winter2019learning} to embed the lead molecule and molecules in the Bank library into continuous implicit space. It is worth noting that the pre-trained encoder has been widely used to perform efficient and smooth search in the implicit space~\cite{QMO,MSO}. 
Next, CMOMO performs linear crossover~\cite{takahashi2001crossover} between the latent vector of lead molecule and that of each molecule in the Bank library, which enables CMOMO to generate a high-quality initial molecular population.  

(2) \textbf{Dynamic cooperative optimization:} 
In this step, the cooperative optimization between the discrete chemical space and the continuous implicit space is dynamically executed in both unconstrained and constrained scenarios.
To be specific, in the unconstrained scenario, CMOMO firstly employs a newly designed vector fragmentation based evolutionary reproduction strategy (VFER) on the implicit molecular population to efficiently generate offspring molecules in the continuous implicit space (Fig.~\ref{framework}B). Next, CMOMO decodes parent and offspring molecules by a pre-trained decoder from the continuous implicit space to the discrete chemical space to evaluate their molecular properties~\cite{winter2019learning} . 
Further, the molecules with better property values are selected by the environmental selection strategy of NSGA-II~\cite{deb2002fast} to obtain the next molecular population. 
CMOMO iterates the above operations until the given number of iterations is reached, which enables CMOMO to possess strong ability to explore the vast search space to obtain molecules with good convergence and diversity, where convergence measures the proximity of molecules to the CPF and diversity assesses the distribution of molecules in the search space.
Afterwards, CMOMO switches to the optimization in the constrained scenario.
In comparison to the optimization in the unconstrained scenario, the optimization within the constrained scenario adopts a novel way (i.e., a ranking aggregation strategy) to evaluate and select molecules by balancing property optimization and constraint satisfaction (Fig.~\ref{framework}C).
Finally, we identify the feasible molecules which have excellent molecular properties and satisfy the drug-like constraints.


\subsection{Experimental design}
To verify the performance of the CMOMO framework, we compare the proposed CMOMO with five state-of-the-art molecular optimization methods on four molecular optimization tasks.
\subsubsection{Optimization tasks.~~}
The four constrained multi-objective molecular optimization tasks contain two benchmark tasks (Task 1 and Task 2) and two practical tasks (Task 3 and Task 4). Their optimization objectives and constraints are described as follows.

\textbf{Optimization objectives.}
Task 1 aims to simultaneously optimize three non-biological activity properties, including the drug likeness (QED)~\cite{QED}, the improvement of the penalized partition ratio of the solute between octanol and water (PlogP\_imp)~\cite{LogP}, and the Tanimoto similarity~\cite{bajusz2015tanimoto} between optimized molecules and lead molecules (Similarity). The three objectives are the larger the better.

Task 2 optimizes three structural scores of a FDA-approved drug (Perindopril)~\cite{brown2019guacamol} and the Similarity. The three structural scores are the dissimilarity score between optimized molecules and Perindopril (Score\_dissmi), the molecular weight score of optimized molecules (Score\_mw), and the rotatable bond score of optimized molecules (Score\_rb). 
The values of the four properties are in the range of $[0,1]$, and the values are the larger the better.

The optimization objectives of Task 3 are QED, the simulated docking score with the $\beta $2-adrenoceptor GPCR receptor (4LDE), and Similarity. The 4LDE protein is responsible for muscle relaxation and bronchodilation~\cite{tartarus}, where the docking scores between molecules and 4LDE protein are simulated by autodock~\cite{eberhardt2021autodock}. 
In the above three optimization objectives, the QED and Similarity are maximization optimization objectives while the 4LDE score is a minimization optimization objective.

The optimization objectives of Task 4 are QED, the predicted GSK3$\beta$ inhibition to glycogen synthase kinase-3 target, the normalized synthetic accessibility (SA), and Similarity. GSK3$\beta $ is known to be implicated in the pathogenesis of several diseases and is a potential target for Alzheimer's disease treatment. In this task, the inhibition of molecules against GSK3$\beta $ target is predicted by a surrogate model~\cite{li2018multi}. All objectives are in the range between 0 and 1, and the values are the larger the better.

\textbf{Constraints.}
Two structural constraints are considered in the above four tasks. 
The first constraint (denoted as ${C_1}$) is that the rings in molecules should have only 5 or 6 atoms, which affects the synthesis of molecules. 
Given a molecule which contains $K$ rings with each ring having ${r_k}$ atoms, its constraint violation on ${C_1}$ is calculated as
\begin{equation}
  {\sigma _1}(x) = \sum\nolimits_{k = 1}^K {[\max \{ {r_k} - 6,0\}  + \max \{ 5 - {r_k},0\} ]}. 
\end{equation}

The second constraint (denoted as ${C_2}$) is that molecules cannot contain toxic or uncommon substructures. Specially, given a molecule $x$, the 163 substructures in MSO~\cite{MSO} (see Figure S1C in the Supplementary Material) are considered to calculate the constraint violation ${\sigma _2}(x) = \left| \xi  \right|$ on ${C_2}$, where $\left| \xi  \right|$ denotes the number of toxic or uncommon substructures in $x$.

Based on ${\sigma _1}(x)$ and ${\sigma _2}(x)$, the aggregated constraint violation degree (\emph{CV}) of molecule $x$ can be calculated as follows.
\begin{equation}
CV(x) = \sum\nolimits_{c = 1}^2 {\frac{1}{{{{\max }_{{x_i} \in \psi }}{\sigma _c}({x_i})}}{\sigma _c}(x)}, 
\end{equation}
where $\psi $ is the set of molecules. 
The larger the value of $CV$, the larger the degree of constraint violation of the molecule.

\subsubsection{Comparison methods.~~}
In the experiments, there are totally five state-of-the-art methods that are selected to compare with the proposed CMOMO, namely QMO\cite{QMO}, Molfinder\cite{MolFinder}, MOMO\cite{xia2022molecule}, MSO\cite{MSO}, and GB-GA-P\cite{GB-GA-NSGAII}. In these five comparison methods, QMO and MSO use aggregation methods to optimize multiple objectives, while the other three methods are Pareto optimization methods. As for the constraint handling strategies of these five methods, QMO, Molfinder, and MOMO deal with constraints by selecting feasible molecules from their final optimization results, MSO treats constraints by aggregating them with optimization objectives, and GB-GA-P addresses constraints by discarding infeasible molecules at each iteration. It is worth noting that, in the experiments, all methods use the same population size/number of samples and the same iterations, which enables the performance comparison to be fair. 

\subsubsection{Evaluation metrics.~~}
We compare the performance of the six methods by using four metrics, namely the optimization success rate (SR), the number of successfully optimized molecules (${N_{SR}}$), the hypervolume (HV), and the mean property value of successfully optimized molecules. Specifically, SR is the ratio of successfully optimized molecules against all molecules to be optimized, where successfully optimized molecules refer to those feasible molecules whose property values are better than the predefined thresholds. The hypervolume ~\cite{while2006faster} is used to comprehensively evaluate both convergence and diversity of successfully optimized molecules, which can be obtained by calculating the size of the hyperspace between the set of molecules and the reference point, where the reference point in each task is set as a zero vector.

In the experiments, the thresholds that are used to assess whether a molecule is successfully optimized are given as follows. For Task 1, QED $\ge 0.85$, PlogP\_imp $\ge 3$, and Similarity $\ge 0.3$;  Task 2, Score\_dissmi$\ge 0.5$, Score\_mw$\ge 0.5$, Score\_rb$\ge 0.5$, and Similarity $\ge 0.3$; Task 3, QED $\ge 0.8$, 4LDE $\le - 10$, and Similarity $\ge 0.3$; Task 4, QED $\ge 0.7$, GSK3$\beta $ inhibition $\ge 0.4$, SA$\ge 0.7$, and Similarity $\ge 0.2$. Besides, since QMO and MSO are aggregation methods that obtain only one molecule in their final results, CMOMO is compared with them only in terms of SR. As for the three Pareto optimization methods, the CMOMO is compared with them using all of the four evaluation metrics. Details about the experimental settings can be found in the Supplementary Material Text S1-S4.

\begin{figure}[h]
 \centering
 \includegraphics[width=0.8\textwidth]{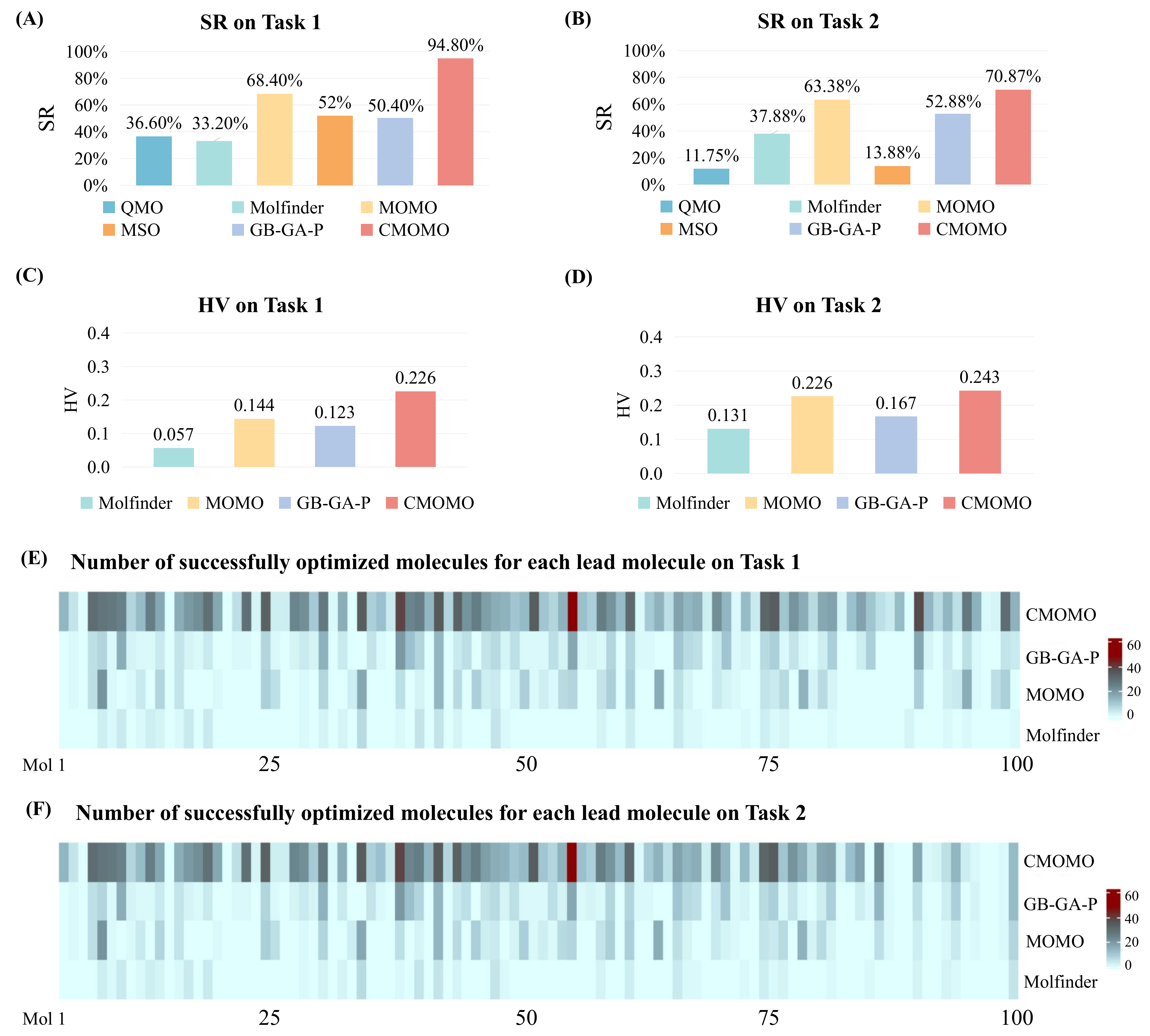}
 \caption{Performance of CMOMO and comparison methods on two benchmark constrained multi-objective optimization tasks. (A) Success rate (SR) of CMOMO and comparison methods on Task 1. (B) SR of CMOMO and comparison methods on Task 2. (C) Hypervolume (HV) of CMOMO and three Pareto optimization methods on Task 1. (D) HV of CMOMO and three Pareto optimization methods on Task 2. (E) The number of successfully optimized molecules obtained by four multi-objective optimization methods on Task 1. (F) The number of successfully optimized molecules obtained by four multi-objective optimization methods on Task 2.}
 \label{Task1,2}
\end{figure}

\subsection{CMOMO outperforms the comparison methods on benchmark tasks}
Fig.~~\ref{Task1,2} shows the success rate, hypervolume, and the number of successfully optimized molecules by the proposed CMOMO and comparison methods.
From the figure, the following three observations can be obtained.

First, the proposed CMOMO achieves the highest success rate in comparison with the five molecule optimization methods on the two benchmark tasks (Fig.~~\ref{Task1,2}A and Fig.~~\ref{Task1,2}B). 
For example, on Task 1, the proposed CMOMO successfully optimized $94.8\%$ of $800$ molecules, which achieves significantly better performance than all the five comparison methods.
It is indicated that the proposed CMOMO is likely to successfully optimize a molecule which is hard to be optimized.


Second, the successfully optimized molecules achieved by the proposed CMOMO have superior molecular properties and exhibit greater diversity, where the diversity can be measured by the number and the distribution of molecules.
(1) To comprehensively access the quality of the successfully optimized molecules, Fig.~~\ref{Task1,2}C and Fig.~~\ref{Task1,2}D present the hypervolume (HV) of the optimized molecules achieved by the proposed CMOMO and other three considered multi-objective methods, i.e., Molfinder, MOMO, and GB-GA-P, where the value of HV measures the convergence and diversity of the solutions.
As can bee seen from the figure, CMOMO obtains the largest mean HV values on the two tasks, which indicates that the proposed CMOMO is capable of generating a wide variety of molecules that have 
superior molecular properties.
(2) To further access the quality of the successfully optimized molecules, we plotted the mean property values of successfully optimized molecules in Figures S2A and S2B in the Supplementary Material.
The results indicate that the proposed CMOMO achieves almost the largest property improvement. 
It is indicated that CMOMO shows the superior performance in enhancing multiple properties of molecules.
Moreover, Figure S4 in the Supplementary Material shows the heatmaps of optimization results obtained by the six methods on Task 1, which displays the property distributions of successfully optimized molecules clearly. From Figure S4, we can see that CMOMO obtains more molecules that exhibit good property values than the comparison methods. 
(3) To check the diversity of the resulted molecules corresponding to each molecule to be optimized,
Fig.~~\ref{Task1,2}E and Fig.~~\ref{Task1,2}F present the number of successfully optimized molecules obtained by four multi-objective optimization methods (CMOMO, GB-GA-P, MOMO, and Molfinder) for $100$ lead molecules.
In our work, the successfully optimized molecules for all lead molecules are also given in Figures S3A and S3B in the Supplementary Material.
We can see from the figure that CMOMO achieved the largest number of successfully optimized molecules for each molecule.
For example, CMOMO obtains $942$ successfully optimized molecules with their QED values being larger than $0.94$, by contrast, only $81$ molecules with QED values larger than $0.94$ are obtained by the best comparison method.

From the above empirical results, we can conclude that the proposed CMOMO is a promising constrained multi-property molecular optimization method.
Besides, the way to deal with constraints by balancing them with property optimization is superior to other existing constraint-handling strategies (e.g., merely selecting feasible candidates from the final optimization outcomes or discarding those deemed infeasible).


\subsection{CMOMO exhibits good performance on finding potential ligands for the 4LDE protein}
\begin{figure}[!t]
 \centering
 \includegraphics[width=0.65\textwidth]{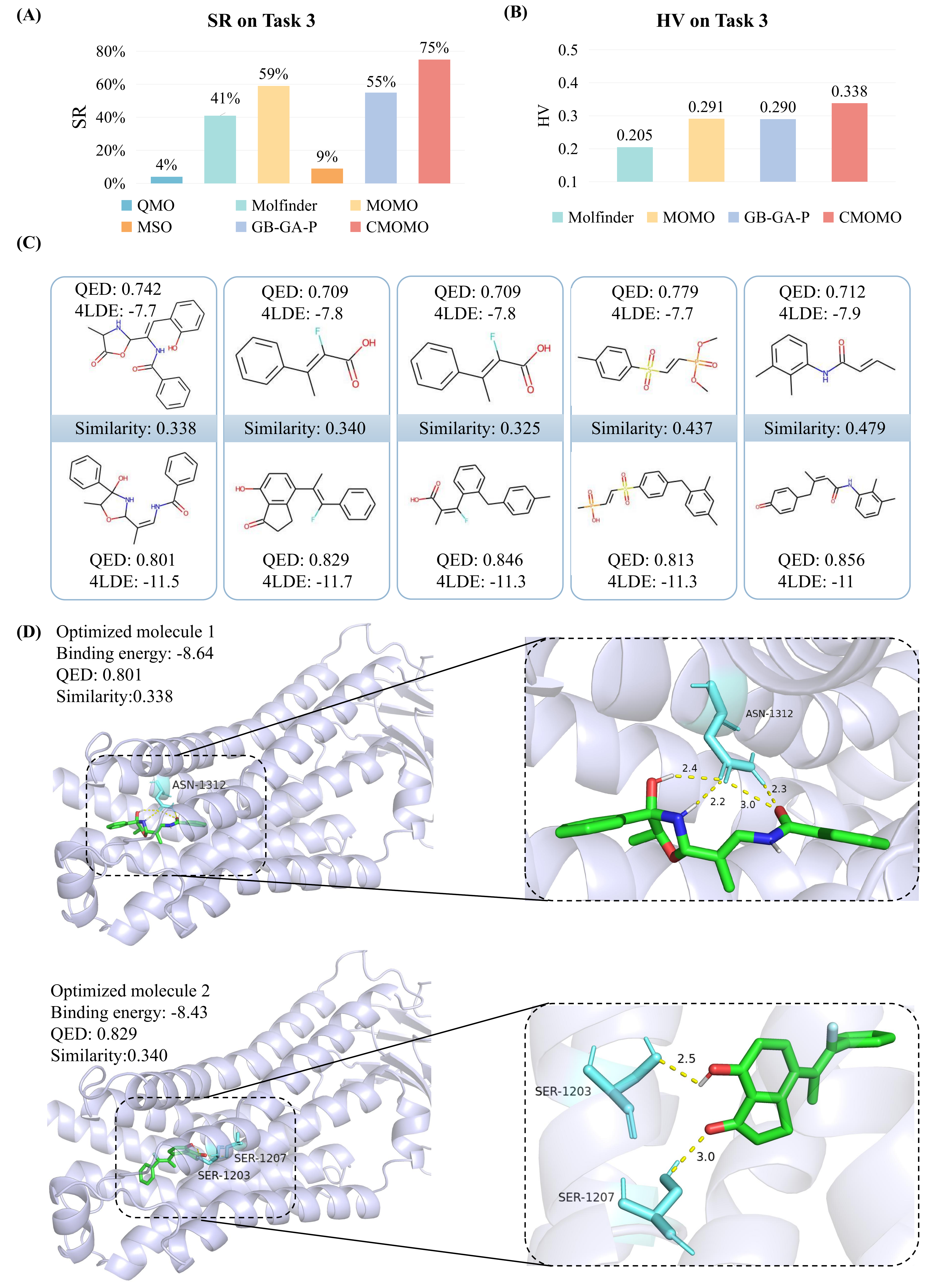}
 \caption{Performance of CMOMO and comparison methods on Task 3. (A) Success rate (SR) of CMOMO and comparison methods on Task 3. (B) Hypervolume (HV) of CMOMO and three Pareto optimization methods on Task 3. (C) The optimization results obtained by CMOMO on five test instances,  where the molecules at the top and bottom rows denote the lead and optimized molecules, respectively. (D) The docking pose, QED, binding energy, and Similarity of two optimized molecules for the 4LDE protein are shown. Both molecules exhibit the desired QED, binding energy, and Similarity. The right two sub-figures show the interactions between the optimized molecules and the amino acid residues in protein binding pockets.}
 \label{Task3}
\end{figure}

In this subsection, we test the performance of CMOMO and comparison methods on the protein-ligand optimization task. First, we give the results of the six considered methods in terms of success rate. As shown in Fig.~\ref{Task3}A, the proposed CMOMO obtains the largest success rate of 75\%, while the largest success rate obtained by the comparison methods is 59\%, which is obviously smaller than the proposed CMOMO. Then, we present their optimization results in terms of HV of successfully optimized molecules, which are shown in Fig.~\ref{Task3}B, the results show that the proposed CMOMO obtains the largest HV value, i.e., 0.338, which is larger than the largest HV value obtained by the other three Pareto optimization methods, i.e., 0.291. The superiority of CMOMO in terms of HV indicates that its successfully optimized molecules have better convergence and diversity. Next, we compare the performance of CMOMO and comparison methods by comparing the quality and quantity of their successfully optimized molecules, where the results are given in Figure S2C and Figure S3C in the Supplementary Material. The results in Figure S2C show that the molecules obtained by CMOMO have better properties. The results in Figure S3C show that CMOMO obtains a larger number of successfully optimized molecules than the three Pareto optimization based comparison methods for most lead molecules.

To better validate the performance of the proposed CMOMO, we provide five pairs of molecules with each pair containing a lead molecule and an optimized molecule to compare their difference. As shown in Fig.~\ref{Task3}C, it can be seen that the proposed CMOMO improves their QED and 4LDE considerably; besides, for each pair of molecules, the optimized and lead molecules are similar, which possess a similarity value that is larger than 0.3. To further verify the performance of our proposed CMOMO, the first two molecules shown in Fig.~\ref{Task3}C are used to conduct the protein-ligand interaction analysis, where the docking software Autodock and the visualization software PyMol~\cite{kagami2020geo} are used to simulate the docking energies and the top docking poses. As shown in Fig.~\ref{Task3}D, the binding energies of the two successfully optimized molecules for the 4LDE protein are smaller than -7 kcal/mol, where the -7 kcal/mol has been widely used as a threshold to assess whether a molecule is drug-like in some researches~\cite{li2023knowledge,ahmad2021molecular}; the low binding energies suggest that these molecules hold potential as ligands for the 4LDE protein. Moreover, the protein-ligand docking poses and the interactions between the two molecules and the 4LDE protein are visualized in Fig.~\ref{Task3}D. It can be seen that the two successfully optimized molecules form 4 and 2 hydrophobic interactions (represented by yellow dashed lines) with the amino acid residues in the binding pocket of 4LDE, respectively. 

Overall, the above statistical results in terms of four evaluation metrics show that the proposed CMOMO outperforms the comparison methods to find potential 4LDE protein ligands; besides, the protein-ligand interaction analysis results show that the successfully optimized molecules obtained by CMOMO have the potential to bind with the 4LDE protein.

\begin{figure}[!h]
 \centering
 \includegraphics[width=0.7\textwidth]{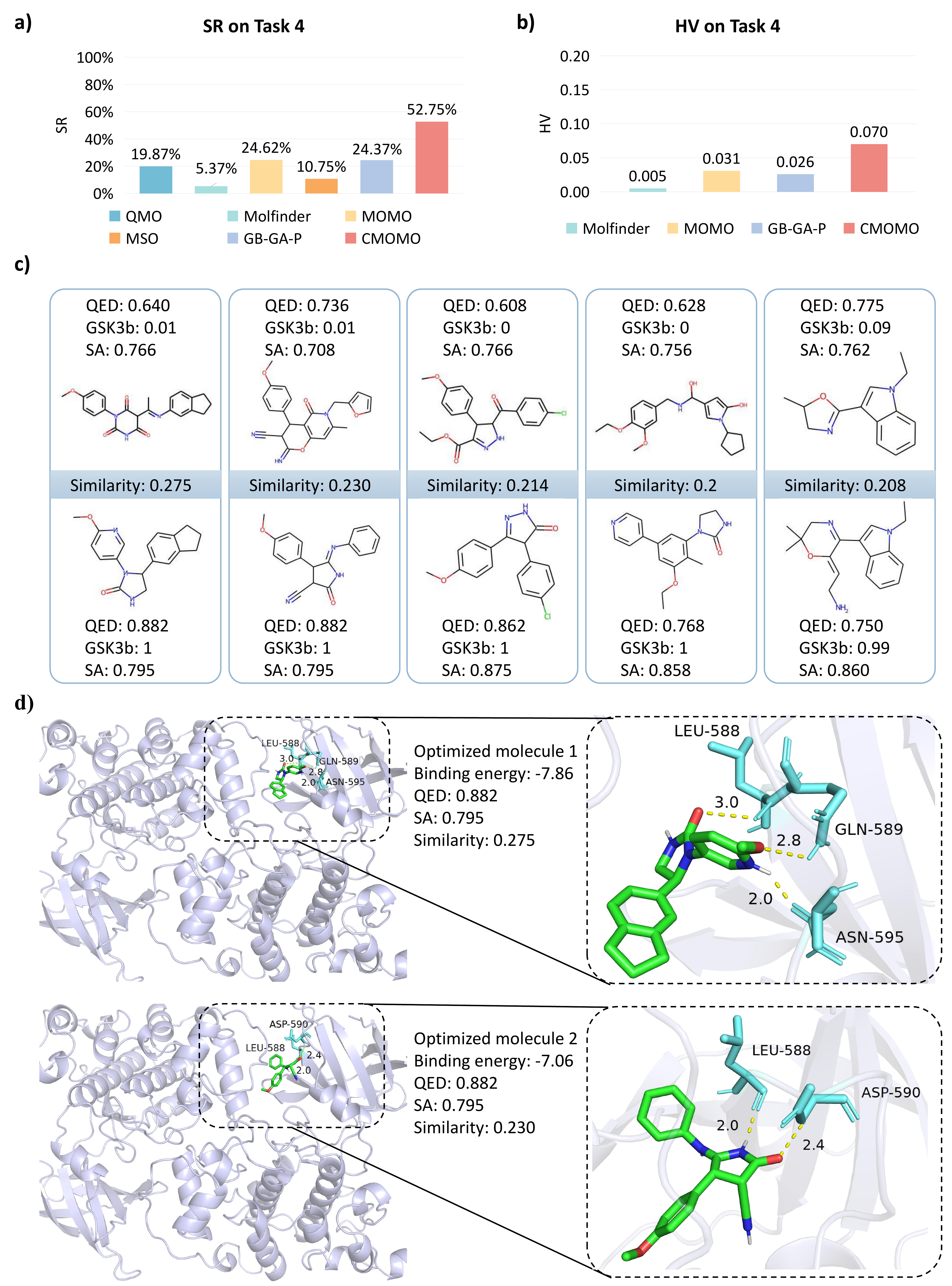}
 \caption{Performance of CMOMO and comparison methods on Task 4. (A) Success rate (SR) of CMOMO and comparison methods on Task 4. (B) Hypervolume (HV) of CMOMO and three Pareto optimization methods on Task 4. (C) The optimization results obtained by CMOMO on five test instances, where the molecules at the top and bottom rows denote the lead and optimized molecules, respectively. (D) The docking poses, QED, binding energy, SA, and Similarity of two optimized molecules for the GSK3$\beta $ target are shown. Both molecules exhibit the desired QED, binding energy, SA, and Similarity. The right two sub-figures show the interactions between the optimized molecules and the amino acid residues in protein binding pockets.}
 \label{Task4}
\end{figure}

\subsection{CMOMO performs well on finding potential inhibitors against the GSK3$\beta $ target}
Fig.~\ref{Task4} presents the results on Task 4, i.e., the inhibitor optimization task, which shows that the proposed CMOMO holds the ability to find potential inhibitors against GSK3$\beta$. To be specific, the proposed CMOMO outperforms the comparison methods in terms of four evaluation metrics, including the success rate, HV, mean property value, and the number of successfully optimized molecules. Besides, the protein-ligand interaction analysis shows that the molecules obtained by CMOMO have potential inhibitions against the GSK3$\beta $ target.

As shown in Fig.~\ref{Task4}A, while the success rates obtained by the five comparison methods range from 5.37\% to 24.6\%, CMOMO achieves the largest success rate of 52.7\%, which is obviously larger than that obtained by the comparison methods. In terms of HV, Fig.~\ref{Task4}B shows that CMOMO obtains the largest mean hypervolume, which is larger than that obtained by three Pareto optimization methods. As for the quality and quantity of successfully optimized molecules obtained by CMOMO and comparison methods, they are given in Figure S2D and Figure S3D in the Supplementary Material. The results in Figure S2D show that the successfully optimized molecules obtained by CMOMO have the best performance on multiple properties, especially in terms of GSK3$\beta $ inhibition and SA. The results in Figure S3D show that CMOMO obtains a larger number of successfully optimized molecules for most lead molecules than comparison methods.

To better validate the performance of the proposed CMOMO, we use five successfully optimized molecules to compare with their corresponding lead molecules. As shown in Fig.~\ref{Task4}C, for each lead molecule, the proposed CMOMO obtains a successfully optimized molecule with the GSK3$\beta $ inhibition, QED and SA being improved considerably. Moreover, we conduct an in-depth analysis of protein-ligand interactions for the first two optimized molecules in Fig.~\ref{Task4}C, where the protein structure of GSK3$\beta$ is downloaded from the UniProt website~\cite{uniprot2019uniprot}. As depicted in Fig.~\ref{Task4}D, the binding energies of the two optimized molecules are also smaller than -7 kcal/mol, when the two molecules generate 3 and 2 hydrophobic interactions (represented by yellow dashed lines) with the amino acid residues of GSK3$\beta$, respectively. The above results illustrate that the molecules obtained by CMOMO have the potential to bind with the GSK3$\beta $ target.

\begin{figure}[!h]
 \centering
 \includegraphics[width=0.5\textwidth]{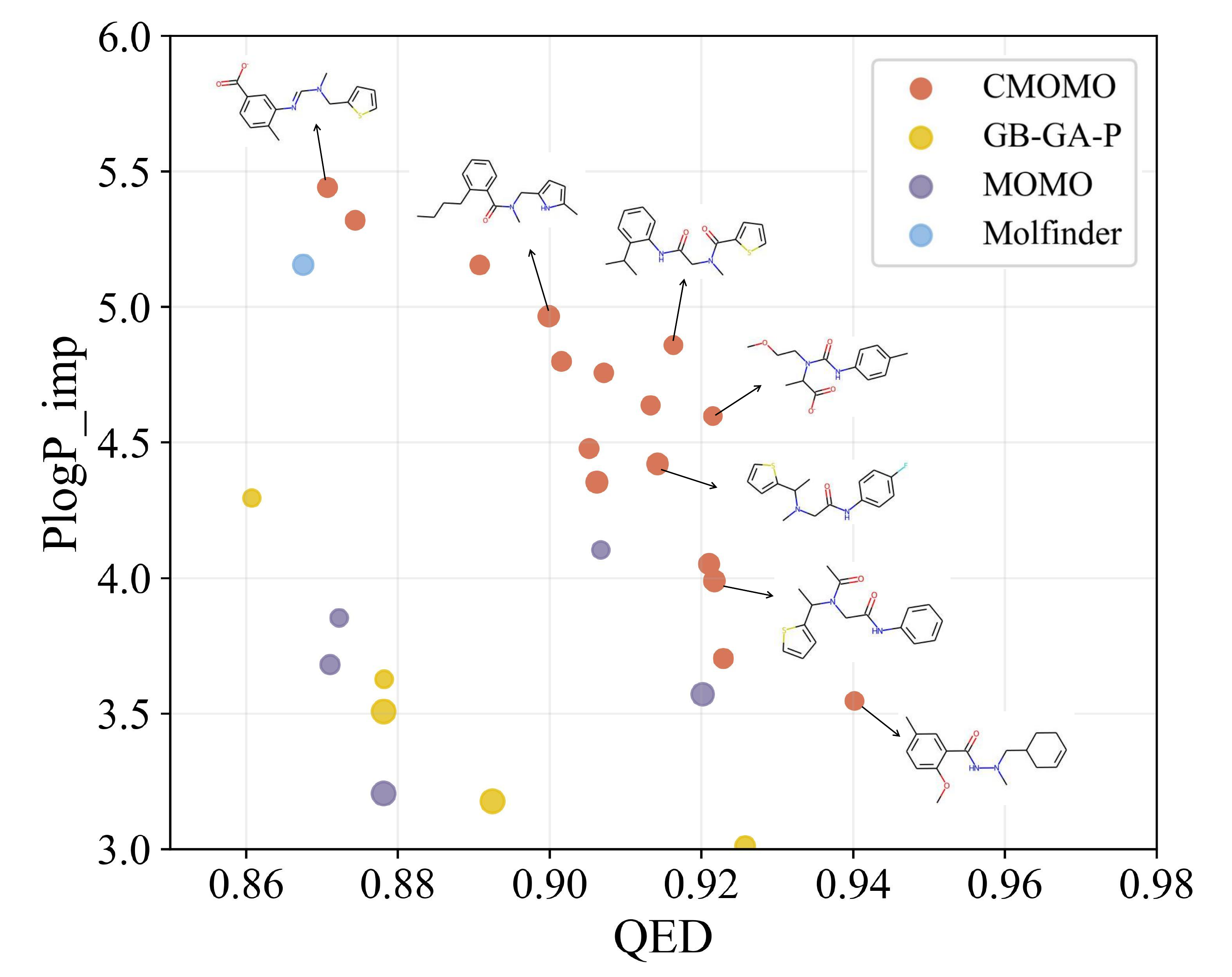}
 \caption{Comparison of successfully optimized molecules obtained by CMOMO and three Pareto optimization methods for a lead molecule. The structures of some molecules obtained by CMOMO are displayed next to the corresponding dots.}
 \label{case1}
\end{figure}

\begin{figure}[!h]
 \centering
 \includegraphics[width=0.9\textwidth]{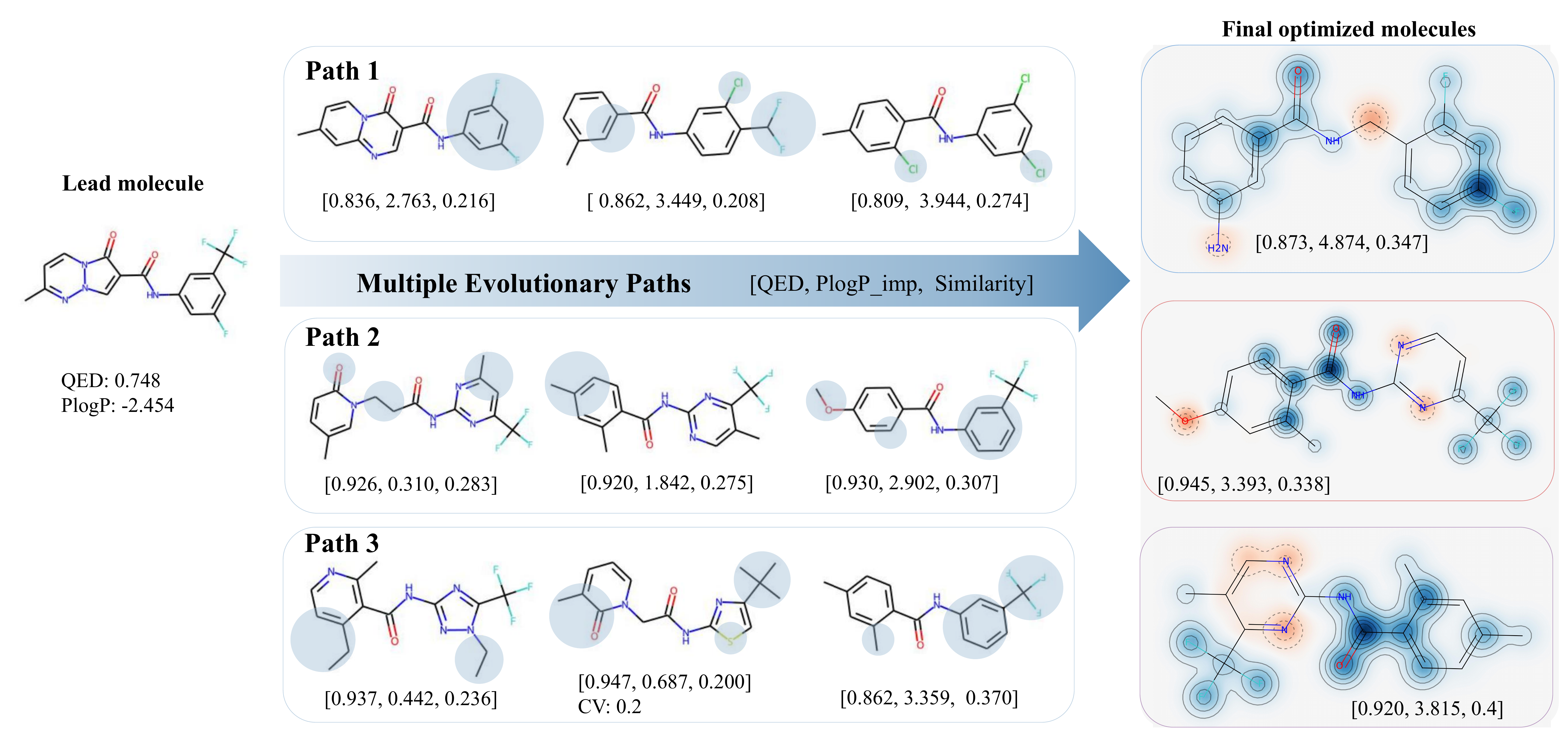}
 \caption{Multiple evolutionary paths formed by the optimized molecules of CMOMO for a lead molecule. The figure consists of three rows showing the evolutionary paths from the lead molecule to the final optimized molecules. The shaded regions in each row highlight the modified substructures. The three Tanimoto similarity maps in the last column show the similarity between the final optimized molecules and the lead molecule, where similar regions are indicated in blue and dissimilar regions are indicated in red.}
 \label{case2}
\end{figure}

\subsection{Case study}
In this subsection, we perform a case study by randomly selecting a lead molecule in Task 1 to investigate the search capability of the proposed CMOMO and the evolutionary trajectory of optimized molecules. Fig.~\ref{case1} gives the optimization results obtained by the proposed CMOMO and the three Pareto optimization based comparison methods, where each dot represents a successfully optimized molecule with its size reflecting the similarity to the lead molecule. It can be seen that the number of successfully optimized molecules obtained by CMOMO is obviously larger than that obtained by the three comparison methods. Besides, the successfully optimized molecules obtained by CMOMO dominate the ones obtained by the comparison methods, which means that the molecules obtained by CMOMO have better QED and PlogP properties. Moreover, compared with the three Pareto optimization based comparison methods, the proposed CMOMO shows competitive performance in terms of the similarity between optimized molecules and the lead molecule. 

Fig.~\ref{case2} presents the evolution trajectories of molecules obtained by CMOMO for a randomly selected lead molecule, where three evolutionary paths are shown. The blue shaded regions in each path denote the modified substructures, the final optimized molecules are provided by plotting three Tanimoto similarity maps in the last column, where similar and dissimilar regions in comparison to the lead molecule are indicated by blue and red colors, respectively. From the figure, it can be seen that the lead molecule can evolve through diverse paths to obtain different successfully optimized molecules that make good trade off between different properties.
To be specific, a final optimized molecule with a large PlogP\_imp value is obtained from the first evolutionary path, the second evolutionary path enables to obtain an optimized molecule with a large QED value, and a final optimized molecule with relatively higher Similarity is obtained from the third evolutionary path. Besides, as shown in Path 3, some infeasible molecules that violate constraints can also be saved in the evolutionary process, which enables the lead molecule to be optimized through multiple evolutionary paths; thus, improving the exploration ability of the proposed CMOMO.

\begin{figure}[h]
 \centering
 \includegraphics[width=0.5\textwidth]{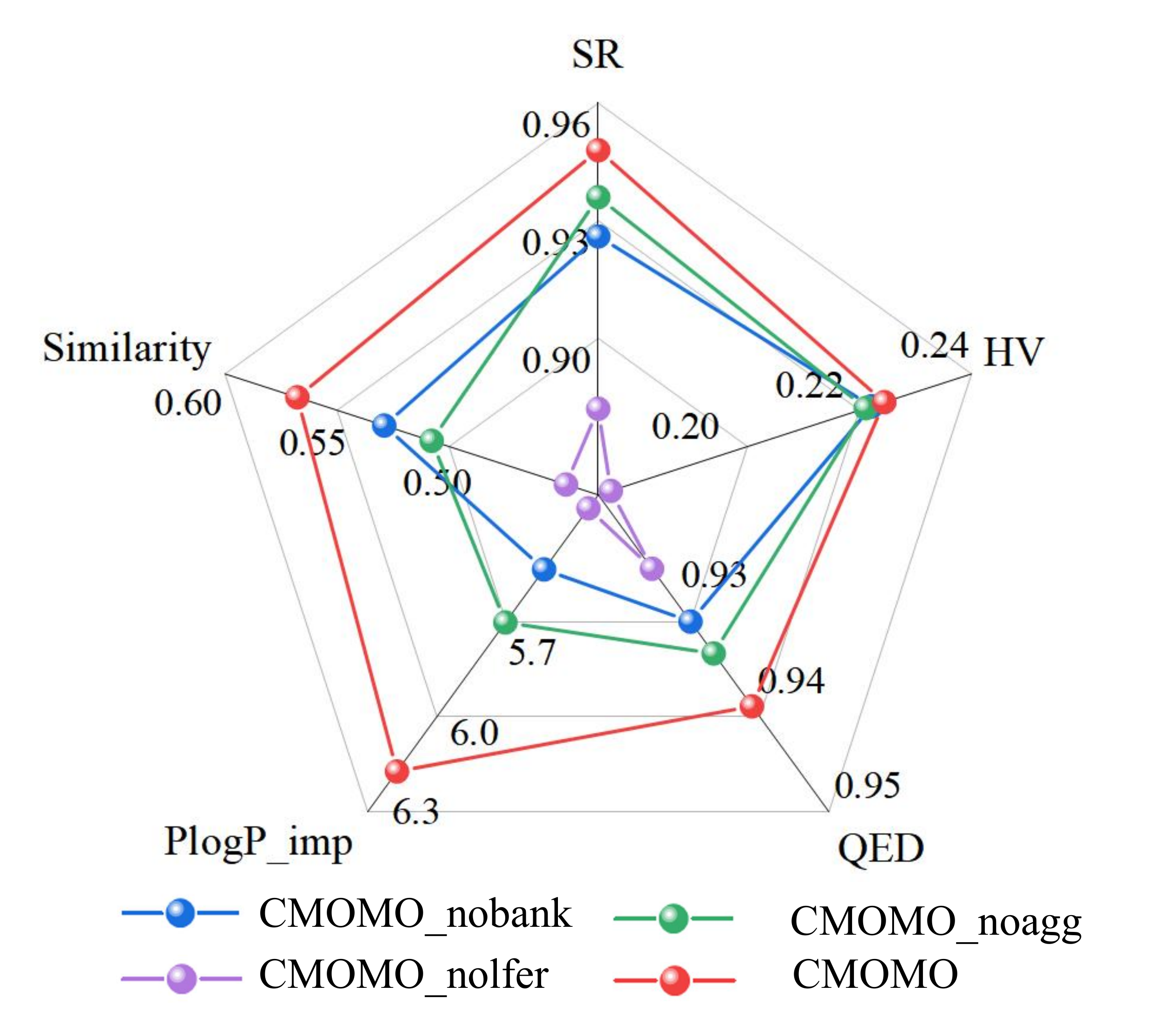}
 \caption{The performance of CMOMO and its three variants on Task 1. The first variant CMOMO\_nobank is obtained by removing the Bank library from population initialization, the second variant CMOMO\_noagg is obtained by replacing the ranking aggregation strategy in the second optimization stage with the constraint dominance principle, the third variant CMOMO\_novfer is attained by replacing the latent vector fragmentation based evolutionary reproduction strategy with the polynomial mutation operation.}
 \label{ablation}
\end{figure}

\subsection{Ablation studies of CMOMO}
\label{sec:ablation}
In this subsection, we conduct some ablation experiments to validate the effectiveness of key components in the proposed CMOMO by designing three algorithmic variants, namely CMOMO\_nobank, CMOMO\_noagg, and CMOMO\_novfer. Specifically, the variant CMOMO\_nobank is obtained by removing the Bank library from the population initialization of CMOMO, the variant CMOMO\_noagg is obtained by replacing the ranking aggregation strategy in the constrained optimization stage of CMOMO with the constraint dominance principle~\cite{deb2002fast}, which is widely used as a constraint handling technique to address constraints, and the variant CMOMO\_novfer is designed by replacing the vector fragmentation based evolutionary reproduction strategy in CMOMO with the polynomial mutation operation, which is widely used by many methods~\cite{DEL,xia2022molecule} to generate the latent vectors of offspring molecules.   

Fig.~\ref{ablation} provides the results of performance comparisons between CMOMO and its three variants on Task 1. It can be seen that CMOMO outperforms the three variants in terms of SR, HV, and three property values, including QED, PlogP\_imp, and Similarity, which verifies that the three components in CMOMO are effective to obtain good performance. Moreover, it can be seen that the performance of CMOMO\_novfer is obviously inferior to CMOMO when the vector fragmentation based evolutionary reproduction strategy is removed from CMOMO, which means that this evolutionary reproduction strategy has the most important effect on the performance of the proposed CMOMO. Although the effects of the Bank library and the ranking aggregation strategy on the performance of CMOMO are not as obvious as that of the latent vector fragmentation based evolutionary reproduction strategy, their effectiveness are also not negligible. For example, when the Bank library is removed from the population initialization of CMOMO to obtain the variant CMOMO\_nobank, the performance of CMOMO on QED and PlogP\_imp degrades considerably, which can also be verified by the violin plots and histogram plots (Figure S5 in the Supplementary Material) of the property distribution of molecules in the initial population.

\section{Discussion}
In this paper, we have proposed CMOMO, a deep constrained multi-objective optimization framework that can be readily adapted to any stringent drug-like criteria and any evaluation metrics of molecular property.
It features an efficient dynamic cooperative optimization that enables CMOMO to balance property optimization and constraint satisfaction.
The proposed CMOMO is able to simultaneously optimize multiple molecular properties while satisfying several stringent drug-like criteria.
Further, the CMOMO framework can be applied to the tasks with other kinds of constraints, and other types of drugs, such as peptide, and proteins.

CMOMO demonstrates its superior performance over baseline results on the simpler benchmark tasks for optimizing two non-biological activity properties (i.e., drug-likeness and penalized logP scores) and similarity or optimizing three structural scores of a drug and similarity, while adhering two constraints (i.e., the rings in molecules should have only 5 or 6 atoms and molecules cannot contain 163 toxic or uncommon substructures).
Under above two constraints, we also apply CMOMO to improve the simulated docking score of existing ligands with the $\beta$2-adrenoceptor GPCR, as well as drug likeness and similarity; and to improve the inhibition of molecules against glycogen synthase kinase-3$\beta$, as well as drug likeness, the normalized synthetic accessibility, and similarity.
The CMOMO-optimized molecules of existing drug molecules show a favourable docking score with $\beta$2-adrenoceptor GPCR and drug likeness score, and satisfy constraints.
Besides, the CMOMO-optimized molecules are consistently predicted to inhibit against glycogen synthase kinase-3$\beta$ and easily be synthesized by property predictors. 
Compared to the five competitors, the proposed CMOMO obtains more successfully optimized feasible molecules with better property values, which verifies the superiority of balancing property optimization and constraint satisfaction.
The multiple evolutionary paths analysis provides insight into how CMOMO efficiently traverses the molecular space to discover a diverse set of improved molecules that possess the desired properties while adhering to established constraints.
The results show strong evidence that CMOMO can serve as a novel and practical tool for molecule optimization to accelerate drug discovery with constraints. 
We also conduct some ablation experiments to validate the effectiveness of the key strategies in the proposed framework, including the population initialization, the vector fragmentation based evolutionary reproduction strategy, and the ranking aggregation strategy.
The results show that CMOMO is an efficient deep constrained multi-objective optimization framework, and the strategy that dynamically handles the constraints across unconstrained and constrained scenarios is superior to existing constraint-handling strategies.

Future work will include taking more and different types of constraints into consideration, and using CMOMO for accelerating the discovery of valid, novel, and high-quality molecules and peptides.
Further, the search effectiveness and efficiency will be further improved by training surrogate models or property predictors, which enables the latent vectors to be optimized multiple times in each iteration.

\section{Materials and Methods}
The multi-property molecular optimization with complex constraints are formulated as constrained multi-objective optimization problems. The proposed CMOMO starts with an effective population initialization, which is used to generate a set of high-quality initial molecules. Then, CMOMO performs dynamic cooperative optimization, where molecules are dynamically optimized by the considered properties and constraints, and cooperatively evolved between discrete chemical and continuous implicit spaces. Specifically, CMOMO employs a dynamic constraint handling strategy to divide the optimization into two scenarios, i.e., unconstrained scenario and constrained scenario, which enables CMOMO to dynamically balance property optimization and constraint satisfaction. In the two scenarios, CMOMO performs the same operation in terms of molecule generation, i.e., the vector fragmentation-based evolutionary reproduction (VFER) strategy in the continuous implicit space, however, they perform different operations in terms of molecule evaluation and selection to present a dynamic constraint handling strategy.

\subsection{Molecular optimization formulation}
In the paper, multi-property molecular optimization problems are formally modeled as constrained multi-objective optimization problems, where each property to be optimized is treated as an optimization objective and the strict requirements are treated as constraints. The proposed constrained multi-objective optimization problems can be mathematically expressed as follows.
\begin{equation}
\begin{array}{ll}\text { Minimize } & \mathbf{F}(x)=\left(f_1(x), \cdots, f_m(x)\right) \\ \text { subject to } & x \in \Omega \\ & g_i(x) \leq 0, \quad i=1, \cdots, p \\ & h_j(x)=0, \quad j=1, \cdots, q\end{array}
\label{equ:ProblemModel}
\end{equation}
where $x$ represents a molecule and $\Omega$ represents the molecular search space. $\mathbf{F}(x)$ is the objective vector consisting of $m$ optimization properties, i.e., $f_1(x), \cdots, f_m(x)$. The $g_i(x)$ and $h_j(x)$ are the $i$-th inequality constraint and the $j$-th equality constraint, respectively.

In the paper we use the constraint violation aggregation function to measure the constraint violation degree of a molecule~\cite{zhang2023multigranularity,zhang2023design}, which is defined as follows.
\begin{equation}
CV(x)=\sum_{i=1}^p \text{max}{(g_i(x),0)} + \sum_{j=1}^q |h_j(x)|
\label{equ:CV}
\end{equation}
If $CV(x)$ is equal to zero, the molecule $x$ is said to be feasible; otherwise, it is an infeasible molecule.

As can be seen from Eq.~\ref{equ:ProblemModel}, the proposed problem model is flexible and scalable, where the optimization objectives can be comprehensive evaluation metrics, non-biological active properties, and biological active properties~\cite{thomas2024molscore,hirlekar2023overview}. As for the constraints, they can be the constraints concerning the ring size, substructure constraints, skeleton constraints~\cite{baell2010new,yet2018privileged}, and many others.

\subsection{Population initialization strategy}
In the proposed CMOMO, the population initialization strategy is used to generate a set of high-property molecules that are similar to the lead molecule. To this aim, we first build a Bank library for each lead molecule by screening high-property molecules from public databases~\cite{sterling2015zinc,gaulton2012chembl, kim2016pubchem} (Figure S6A in the Supplementary Material). Then, we use the pre-trained encoder to encode the lead molecule and screened molecules which are represented by SMILES strings into the continuous implicit space. Next, a linear crossover operation is performed between the latent vector of lead molecule and that of each screened molecule to obtain a set of new latent vectors (Figure S6B in the Supplementary Material). Finally, these newly generated latent vectors are decoded from the continuous implicit space to the discrete chemical space. In this way, molecules in the initial population possess not only the genes of lead molecule but also that of screened molecules, which enables initial molecules to have good properties under the premise that they are similar to the lead molecule.

\subsection{Molecule generation strategy}
At the two optimization scenarios, CMOMO utilizes pre-trained encoder and decoder to map molecules between the discrete chemical space and continuous implicit space, which enables it generate molecules effectively as QMO~\cite{QMO} and MSO~\cite{MSO}.
Besides, a vector fragmentation based evolutionary reproduction (VFER) strategy is designed in CMOMO to further enhance the effectiveness and efficiency of generating offspring molecules, which consists of two operations, i.e., blended linear crossover and fragmentation-based mutation. To be specific, CMOMO first selects two latent vectors randomly as parent molecules to perform linear crossover, then, the generated vectors are divided into small fragments with one fragment being selected randomly to perform mutation.

\textbf{Crossover.} In the process of generating offspring molecules, the crossover operation is mainly used to inherit the good genes of parent molecules. To this aim, the blended linear crossover operator~\cite{takahashi2001crossover} is used in CMOMO, which has also been widely used by many other methods, including DEL~\cite{DEL} and MOMO~\cite{xia2022molecule}.
Specifically, given a set of latent vectors of molecules, two vectors (${z_1}$ and ${z_2}$) are randomly selected as parent molecules to generate two offspring molecules (${{z'}_1}$ and ${{z'}_2}$) by the following way.
\begin{equation}
\left\{ \begin{array}{l}
{{z'}_1} = {z_1} + ( - d + (1 + 2d){u_1})({z_2} - {z_1}),\\
{{z'}_2} = {z_1} + ( - d + (1 + 2d){u_2})({z_2} - {z_1}),
\end{array} \right.
\end{equation}
where ${u_1}$ and ${u_2}$ are two random numbers that distribute uniformly between 0 and 1. The $d \ge 0$ is a parameter that controls whether the search space is interpolated or extrapolated by the blended linear crossover operator. 

\begin{figure}[h]
 \centering
 \includegraphics[width=0.35\textwidth]{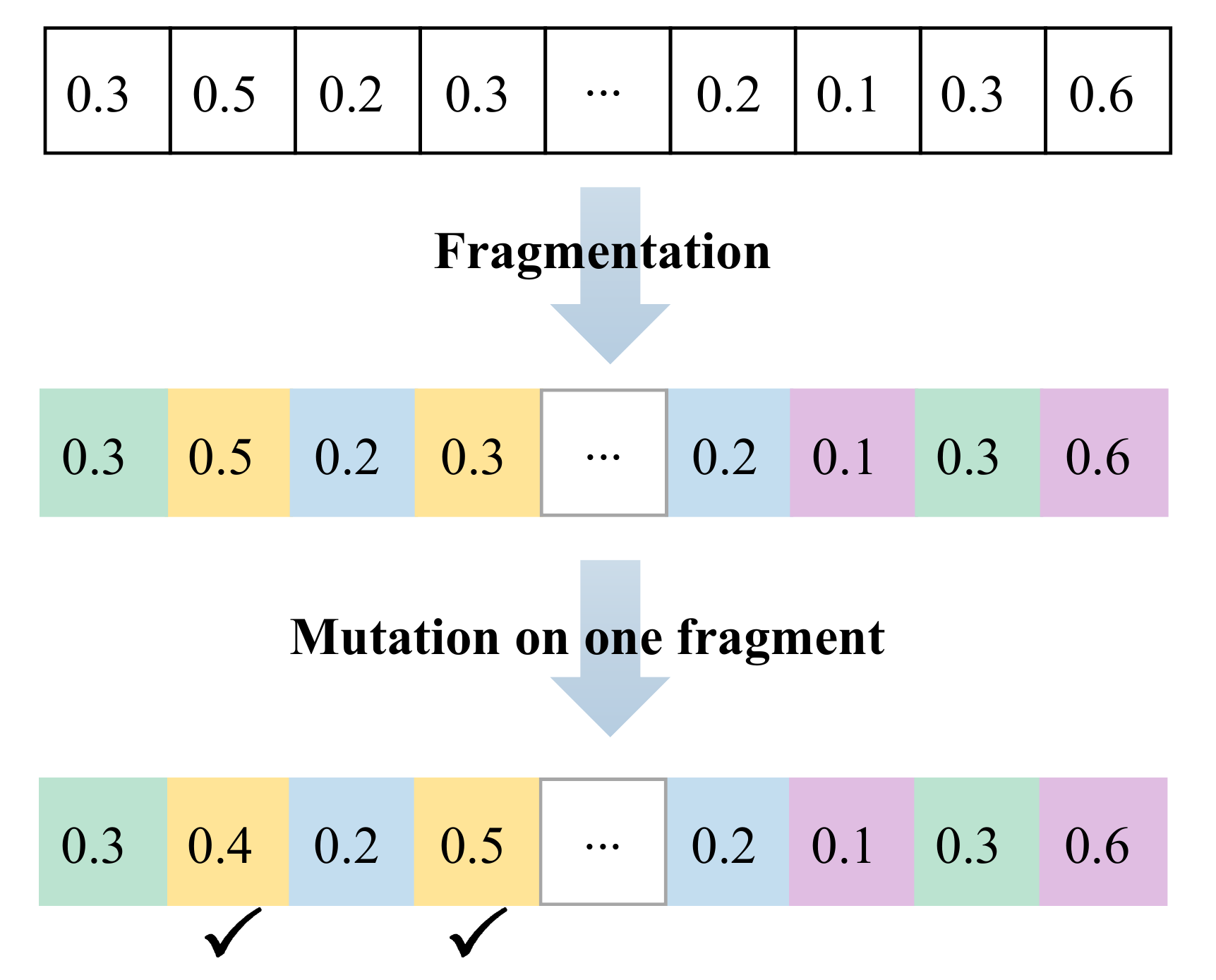}
 \caption{Illustrative example of the proposed fragmentation-based mutation operator. Given a latent vector generated by the crossover operator, the mutation operator first divides it into multiple fragments. Then, one fragment is selected randomly to mutate all genes within it.}
 \label{mutation}
\end{figure}

\textbf{Mutation.} In the proposed algorithm, the mutation operator is used to introduce some new genes, which prevents molecules from trapping into local optima. Besides, since the length of latent vectors can be very large (e..g, 512 in QMO, MSO and the proposed CMOMO), herein a fragmentation-based mutation operator is designed. As shown in Fig.~\ref{mutation}, given a latent vector generated by the crossover operator, the mutation operator first divides it into some small fragments. Then, the mutation operator selects a fragment randomly to mutate all genes in it with the mutation probability ${p_m}$. In this way, the dimension of the long latent vector can be reduced considerably, which alleviates the issue of the curse of dimensionality. Besides, at each time, the operator mutates multiple genes instead of only one gene, which facilitates to improve the search efficiency of the proposed method. It is worth noting that the effectiveness of the proposed fragmentation-based mutation operator has been verified by the ablation study in section~\ref{sec:ablation}.   

\subsection{Dynamic constraint handling strategy}
When CMOMO adopts the same strategy to generate high-quality molecules in the two scenarios, the two optimization scenarios are different in terms of molecule evaluation and selection. Specifically, when the first scenario evaluates and selects molecules by only considering property values, the second scenario takes both property values and constraint violation degree into consideration. In this way, the complex constraints in multi-property molecular optimization are addressed dynamically, which enables the final results obtained by CMOMO to be feasible with good convergence and diversity.

\subsubsection{Molecule evaluation and selection in the first scenario.~~}
\label{sec:MES1}
In the proposed CMOMO, the first scenario is an unconstrained multi-objective molecular optimization scenario, which aims to find molecules with good convergence and diversity. To this aim, when CMOMO in this scenario evaluates molecules by considering only their property values, the environmental selection strategy of NSGA-II is used to select molecules from the union of parent and offspring molecules, which consists of two components, i.e., non-dominating sorting and crowding distance calculation.

\textbf{Non-dominated sorting~\cite{verma2021comprehensive}.} The non-dominated sorting is used to assign molecules to different fronts according to their dominance relationship, which facilitates to improve the convergence of molecular population by selecting molecules with smaller front number. Typically, molecules with smaller front number are the ones with good convergence. Herein, we first introduce how to determine the dominance relationship between different molecules. Given two molecules ${x_1}$ and ${x_2}$, the molecule ${x_1}$ is said to dominate molecule ${x_2}$ (denoted as ${x_1} \prec {x_2}$) if and only if ${x_1}$ is not worse than ${x_2}$ on any objective, besides, the molecule ${x_1}$ is better than ${x_2}$ on at least one objective. Given that the collection of all molecules to be sorted is termed as $Q$, if there does not exist a molecule in $Q$ that dominates molecule ${x_1}$, the molecule ${x_1}$ is called a non-dominated molecule; otherwise, it is called a dominated molecule. In non-dominated sorting, all non-dominated molecules are first selected and removed from $Q$ to obtain the first front ${F_1}$. Then, the dominance relationship between the remaining molecules in $Q$ is recalculated with all non-dominated molecules being removed from $Q$ to obtain the second front ${F_2}$. The above operations repeat until all molecules in $Q$ are assigned to different fronts.   

\textbf{Crowding distance calculation~\cite{verma2021comprehensive}.} In the cast that the non-dominated sorting is used to improve the convergence of molecular population, molecules with the same front number are further sorted by their crowding distance, which helps to maintain the diversity of molecular population. Specifically, given a molecule $x$, its crowding distance is calculated as follows.
\begin{equation}
\label{CD}
CD(x) = \sum\nolimits_{i = 1}^m {\frac{{{f_i}({x_a}) - {f_i}({x_b})}}{{f_i^{max} - f_i^{min}}}}, 
\end{equation}
where ${f_i^{max}}$ and ${f_i^{min}}$ are the maximum and minimum objective values of all molecules on the $i$-th objective, $x_a$ and $x_b$ are two nearest neighborhood molecules to molecule $x$ with their objective values ${{f_i}({x_a})}$ and ${{f_i}({x_b})}$ being larger and smaller than ${{f_i}({x})}$, respectively.

\textbf{Molecule selection.} In the first scenario, when the front number and crowding distance of each molecule are obtained, the proposed CMOMO selects $P$ molecules from the union of parent and offspring molecules that contains $2P$ molecules, where the front number is used as the first selection criterion and the crowding distance is adopted as the second selection criterion. Specifically, CMOMO first selects molecules front by front until the number of molecules in $F_{1} \bigcup F_{2} \bigcup \cdots \bigcup F_{k}$ is larger than $P$. Then, CMOMO deletes $|F_{1} \bigcup F_{2} \bigcup \cdots \bigcup F_{k}| - P$ molecules with the smallest crowding distance from $F_{k}$. In this way, CMOMO obtains $P$ molecules with good convergence and diversity to undergo further optimization.

\subsubsection{Molecule evaluation and selection in the second scenario.~~}
\label{sec:MES2}
Different from the first scenario, the second scenario is a constrained multi-objective molecular optimization scenario, which poses high requirement on the balance between property optimization and constraint satisfaction. To address this issue, when the second scenario considers both property values and constraint violation degree in molecule evaluation, a ranking aggregation strategy is designed to perform molecule selection, which consists of three steps, i.e., property-preference ranking, constraint-preference ranking, and comprehensive scoring (Fig.~\ref{framework}C).

\textbf{Property-preference ranking.}
The property-preference ranking considers only property values of molecules. Given the union of parent and offspring molecules which contains $2P$ molecules, when their front numbers and crowding distances are obtained by non-dominated sorting and crowding distance calculation as introduced in section~\ref{sec:MES1}, these $2P$ molecules are ranked by the following ways. First, the molecules are ranked based on their front numbers, where the molecules having smaller front number are ranked ahead of that having larger front number. Then, the molecules having the same front number are ranked based on their crowding distances, where the molecules having larger crowding distances are ranked ahead of that having smaller crowding distances. In this way, each molecule in the union of parent and offspring molecules obtains a unique rank number ranging from 1 to $2P$. For example, the molecule having the smallest front number and largest crowding distance is given a rank number of 1, by contrast, the molecule having the largest front number and smallest crowding distance is assigned a rank number of $2P$.

\textbf{Constraint-preference ranking.}
The difference between the constraint-preference ranking and property-preference ranking lies in the fact that the constraint-preference ranking takes both property values and constraint violation degree into consideration. When the two ranking operations rank molecules in the union of parent and offspring molecules according to front numbers and crowding distances, they use different methods to obtain the front numbers of molecules. Specifically, in property-preference ranking, front numbers of molecules are obtained by non-dominated sorting as introduced in section~\ref{sec:MES1}, which does not consider constraints. By contrast, in constraint-preference ranking, front numbers of molecules are obtained by the constraint dominance principle. Given two molecules ${x_1}$ and ${x_2}$, the molecule ${x_1}$ is said to dominate molecule ${x_2}$ if one of the following conditions is met: (1) both ${x_1}$ and ${x_2}$ are feasible, besides, ${x_1}$ has better property values than ${x_2}$. (2) ${x_1}$ is feasible while ${x_2}$ is infeasible; (3) both ${x_1}$ and ${x_2}$ are infeasible, besides, the constraint violation degree of ${x_1}$ is smaller than that of ${x_2}$.

\textbf{Comprehensive scoring.}
When the above two ranking operations are performed, each molecule obtains two rank numbers, which are herein used to perform a comprehensive scoring for all molecules. Specifically, the rank numbers obtained in property-preference ranking are saved in $S_1$ with $S_1^{i}$ representing the first score of the $i$-th molecule, while the rank numbers obtained in constraint-preference ranking are saved in $S_2$ with $S_2^{i}$ representing the second score of the $i$-th molecule. In this way, each molecule in the union of parent and offspring molecules obtains a comprehensive score by aggregating its two scores as follows. 
\begin{equation}
\begin{array}{l}
S^{i} = \alpha {S_1^{i}} + (1 - \alpha ){S_2^{i}},\\
\alpha  = \frac{1}{2} \times (1 + \cos (\frac{t}{T}\pi )).
\end{array}
\end{equation}
where the $S^{i}$ represents the comprehensive score of the $i$-th molecule, the $\alpha$ is a parameter which gradually decays from 1 to 0 as the evolutionary generation $t$ increases.

Based on the comprehensive score of each molecule, CMOMO selects $P$ molecules with the smallest comprehensive scores from the union which contains $2P$ parent and offspring molecules. Besides, CMOMO dynamically adjusts the effects of property-preference ranking and constraint-preference ranking on the comprehensive scoring, which enables the algorithm to focus on property optimization and constraint satisfaction respectively at the early and later evolutionary stages; thus, facilitating to obtain feasible molecules with good convergence and diversity.

\section*{Acknowledgments}
\subsection*{Author Contributions} 
Y. S. conceived the study. 
X. X. constructed the databases and developed the codes, X. X., Y. Z., and Y. S. implemented experiments and analyzed the results. X. X., Y. Z., X. Z., X. Z., C. Z., and Y. S. wrote and critically revised the manuscript.

\subsection*{Funding}
This work was funded by the National Key Research and Development Program of China (2021YFE0102100); National Natural Science Foundation of China (62172002, 62202004, 62322301); The University Synergy Innovation Program of Anhui Province(GXXT-2022-035); Anhui Provincial Natural Science Foundation (2108085QF267, 2008085QF294); The University Outstanding Youth Research Project of Anhui Province (2022AH020010); The University Synergy Innovation Program of Anhui Province (No. GXXT-2021-039); The Project of Key Laboratory of Intelligent Computing \& Signal Processing (Anhui University), Ministry of Education(2020A005).

\subsection*{Conflicts of Interest}
The authors declare that there is no conflict of interest regarding the publication of this article.

\subsection*{Data Availability}
The datasets used in this project are updated and available at https://github.com/ahu-bioinf-lab/CMOMO-master. 

\subsection*{Code Availability}
All of the codes are updated and available at https://github.com/ahu-bioinf-lab/CMOMO-master.

\section*{Supplementary Materials}
The supplementary materials contains the details of datasets, comparison methods, evaluation metrics, parameter settings, Bank library for 4 tasks and six supplementary figure. 

A link to access the supplementary materials will be provided in the published article.

\printbibliography


\end{document}